\RequirePackage{fix-cm}

\documentclass[a4paper,11pt]{article}

\usepackage{graphicx}
\usepackage[utf8x]{inputenc}
\usepackage{todonotes}
\usepackage{subfigure}
\usepackage{hyperref}
\usepackage{pgfplots}
\usepackage{url}

\pgfplotsset{table/col sep=semicolon}

\title{An Experimental Evaluation of Machine-to-Machine Coordination Middleware: Extended Version}
\author{Filipe Campos, José Pereira\\
\begin{small}High-Assurance Software Laboratory\end{small}\\
\begin{small}INESC TEC \& University of Minho\end{small}\\
\begin{small}\{fcampos, jop\}@di.uminho.pt\end{small}\\\\\\
Technical Report\\\\\\
}

\date{\today}

\begin{document}

\maketitle
\begin{abstract}
The vision of the Internet-of-Things (IoT) embodies the seamless discovery, configuration, and interoperability of networked devices in various settings, ranging from home automation and multimedia to autonomous vehicles and manufacturing equipment.
As these applications become increasingly critical, the middleware coping with Machine-to-Machine (M2M) communication and coordination has to deal with fault tolerance and increasing complexity, while still abiding to resource constraints of target devices.

In this report, we focus on configuration management and coordination of services in a M2M scenario.
On one hand, we consider ZooKeeper, originally developed for cloud data centers, offering a simple file-system abstraction, and embodying replication for fault-tolerance and scalability based on a consensus protocol.
On the other hand, we consider the Devices Profile for Web Services (DPWS) stack with replicated services based on our implementation of the Raft consensus protocol. We show that the latter offers adequate performance for the targeted applications while providing increasing flexibility.
\end{abstract}




\section{Introduction}

The challenges of interoperability, composability, and long term maintainability in Machine-to-Machine (M2M) applications in a variety of environments have been approaching those of enterprise computing. The first driver for this is that it has become cost-effective to equip most devices with a considerable a{\-}mount of processing and networking resources, making Ethernet networking and mainstream operating systems (e.g. Linux) ubiquitous and increasing the variety and complexity of functions performed. The second driver is the increasing diversity as previously closed systems are opened up to multiple vendors and become full fledged service ecosystems. Finally, these systems play increasingly important roles in day-to-day life, raising more vehemently the issue of dependability.

Service-oriented architectures (SOA) have been a mainstay in enterprise computing, addressing similar needs, and there is now a growing interest in services for systems of connected devices.
The Devices Profile for Web Services (DPWS) addresses this with a set of protocols that resource constrained devices should implement in order to achieve seamless networking and interoperability through Web Services.
It assumes that each device behaves as a standard \emph{hosting service}, providing basal functionality, and exposing one or more \emph{hosted services} that offer device specific functionality.
It includes basic SOAP, WSDL, HTTP binding, WS-Ad\-dress\-ing, and WS-Security that are at the core of Web Services capabilities and interoperability. It also includes WS-E\-vent\-ing, to allow event notification, and SOAP-over-UDP, which enables the usage of UDP as a transport for SOAP messages and enables network level multicast, thus paving the way for dynamic discovery in WS-Discovery\,\cite{ws_disc}, WS-Meta\-data\-Exchange, and WS-Policy.
These protocols allow a client to discover devices in the network, and to learn about their services, resources and characteristics.
Multiple open source implementations of the DPWS exist\,\cite{ws4d_url,soa4d_url}, and the Windows operating systems are shipped with a built-in DPWS framework, thus rendering this specification available in most personal computers and in many devices such as set-top boxes.
Although DPWS provides an adequate infrastructure for small scale systems, such as home automation, it does not provide any fault tolerance protocol for Web Services.

The alternative is resorting to coordination middleware that has been developed in the context of large scale cloud computing infrastructures. For instance, Google's Chubby\,\cite{Burrows:2006:CLS:1298455.1298487} provides the abstraction of lock service and encapsulates a consensus protocol, easing the implementation of distributed fault-tolerant primitives. The most commonly used is
Apache ZooKeeper\,\cite{Hunt:2010p21300}, a highly-available coordination service which provides synchronization and group service. It provides consistent shared state through a simple file-system interface, that allows storing shared information and allows synchronization primitives to be easily implemented by offering sequential counters and exclusive node creation operations. It also supports client failure detection and membership management by means of ephemeral nodes, that are removed when the corresponding session is terminated. Finally, it supports limited event dissemination by notifying watchers when nodes are changed.
Its implementation adopts the replicated state machine approach and it follows the primary-backup scheme, using the ZooKeeper Atomic Broadcast (Zab) protocol\,\cite{Junqueira:2011p21339} to propagate incremental state updates, as a result of handling client requests, from the primary to the backup servers.
ZooKeeper's replicated service can progress while tolerating a minority of failed servers, which can later recover and rejoin the ensemble or cluster.

Briefly, ZooKeeper provides a strong foundation for reliable service coordination, but rests on a heavyweight server model. It requires Java SE 1.6 to run, not being compatible with the Java Micro Edition, which is the Java platform normally available at resource-constrained platforms, or even with the Android operating system, which is increasingly being adopted in such platforms. On the other hand, DPWS provides a flexible platform for service implementation, with generic service and event dissemination capabilities, but falls short in terms of fault-tolerance.
In fact, although conceivable, an adaptation of existing Web Service coordination protocols such WS-AtomicTransaction to provide transaction-based reliability are not desirable, due to their implicit fault model tailored to enterprise infrastructure.

In this report we seek a compromise between these two alternatives by investigating what it takes to provide on DPWS the missing functionality of a coordination service such as ZooKeeper. As DPWS already embodies event notification, this means providing replicated fault-tolerant shared data. Then we compare the resulting performance with ZooKeeper on the target scenarios, with limited resources and smaller scale deployments.  The core of our approach is to implement the Raft\,\cite{Ongaro:2014p20967} consensus protocol on the WS4D stack to build replicated Web Services that are able to tolerate both network and server failures. This implementation leverages existing DPWS components such as WS-Discovery for cluster membership maintenance and can easily be combined with WS-Eventing for notification.

The rest of this report is organized as follows: Section\,\ref{sec:bg} provides background knowledge on existing dependability standards for Web Services, on the consensus problem and related algorithms. Section\,\ref{sec:proposal} describes the components of the proposed and some application scenarios. Section\,\ref{sec:results} describes the experimental settings and comments on the obtained results. Section\,\ref{sec:related} presents and discusses some of the related work, while Section\,\ref{sec:conclusion} finalizes the report with some concluding remarks and provides directions for future work.
\section{Background}
\label{sec:bg}

Aiming to bridge the gap between what is provided by DPWS and ZooKeeper in terms of support for dependable services, we first describe the common approach for service coordination in enterprise systems with Web Services. As these mechanisms do not fit our requirements, we then describe the consensus approach at the core of ZooKeeper, assessing its fitness for our purpose.

\subsection{Web Services Standards}

The term coordination sometimes refers to a type of orchestration defined in WS-Coordination, which specifies an extensible framework for context management, that provides coordination for the actions of distributed applications.
This coordination is achieved through provided protocols that support distributed applications, for instance, those that need to reach consistent agreement on the outcome of distributed transactions.
Both WS-AtomicTransaction (WS-AT) and WS-BusinessActivity (WS-BA) extend WS-Coordination by defining their own coordination type: short-term atomic transactions, and long-running business activities, respectively.
Focussing on WS-AT, it provides an adaptation for Web Services of the classic \textit{2PC} mechanism, which is often said that does not adapt well to Web Services\,\cite{understanding_soa}.
Nonetheless, it is adequate for interoperability across short-lived, co-located services that need to ensure consistent, all-or-nothing results for a transaction.
As mentioned in the previous section, transaction-based reliability, albeit capable of dealing with server failures, is not always desired, specially in scenarios with resource-constrained devices, which sets asides this type of standards.

WS-ReliableMessaging (WS-RM) can be combined with several WS-* standards.
On the one hand, it can improve its features by leveraging WS-Addressing, for identifying messages and endpoints, WS-Security, to protect the integrity and confidentiality of the exchanged messages, and WS-Policy, to specify the delivery assurance, among other sequence requirements\,\cite{soa_concepts_tech_design}.
On the other hand, due to its ability to ensure reliable communication between two endpoints, WS-RM can be leveraged by other standards, such as WS-Eventing, WS-Notification, WS-AT, WS-BA and WS-Coordination to achieve reliable communication among the intervening parties.
Although the WS-RM specification allows to condition service activities, it is different from WS-AT or WS-BA, in the sense that a coordinating entity is not needed to inspect the progress of the activities, being the reliability rules conveyed as SOAP headers in the exchanged messages\,\cite{soa_concepts_tech_design}.
WS-RM would be a suitable standard to ensure point-to-point reliable message delivery.
However, it would be very inefficient and poise a heavy weight on the message sender in terms of processing power, if there are lots of message recipients or if lots of errors occur.
In order for WS-RM to guarantee atomic delivery to all targets, it would have to rely on WS-AT, or a similar coordination protocol, which would increase furthermore the consumption of the sender's processing and communication resources, due to the additional message traffic.
Hence, WS-RM, by itself, is not capable of dealing with failures in a cluster that provides a replicated service.
This emphasizes the need of a capable, albeit resource-efficient, algorithm for such a scenario.

\subsection{Consensus}

Consensus is an abstraction of the problem of all processes in a fault-tolerant distributed system agreeing on the same value despite having started with different opinions and regardless of some of them failing\,\cite{1353023}.
Consensus protocols are the basis for the state machine approach to distributed computing\,\cite{359563}.
This technique allows the conversion of an algorithm into a fault-tolerant and distributed implementation, through the ordering of all the actions involved.
This ordering mechanism depends on the synchronization of the actions among all the processes or nodes involved in the distributed system.
To achieve a consistent order in all the nodes, a consensus protocol is essential.
This approach\,\cite{359563} also handles safely all cases of failure, since failure can only be perceived in the context of physical time, by a user or a process if a supposedly failed process is taking too long to respond.
Consensus algorithms for practical systems can progress correctly as long as a majority of the servers haven't failed and if they can communicate with each other and with clients.


Raft\,\cite{Ongaro:2014p20967} is a consensus algorithm that follows the replicated state machine approach.
It decouples key elements of consensus, like leader election, log replication and safety, while enforcing a stronger degree of coherency to reduce the number of possible states.
Raft is similar to other consensus algorithms, such as Viewstamped Replication\,\cite{Oki:1988p21387,Liskov:2012p21386}, but it stands out due to its strong leadership, as the leader concentrates as much functionality as possible, and as the leader election is fundamental for its consensus protocol.
Another key feature of Raft is its mechanism to support cluster membership changes, where the majorities of two different configurations overlap, allowing the cluster to operate normally during such transitions.

A Raft cluster is composed by several servers, being five a typical setup, which allows the system to tolerate two failures.
Each server is always in one of three states: leader, follower or candidate; and a cluster in normal operation contains a single leader, being the remaining servers in the follower state.
The leader serves client requests and controls their application to the replicated log and state machine.
The candidate state is the transient state from follower to leader, during which the server starts an election trying to be elected as the new leader of the cluster.
The follower state is passive, which means it simply responds to invocations from the leader and candidates.
Raft uses the notion of term as an arbitrary period of time that starts with an election, where one or more candidates try to become the leader, but where at most one can take that role.
If a candidate succeeds and becomes the new leader, it will keep that role until a new term is started.
If there is no winner, a new election will be started, consequently on a new term.
Terms are numbered with consecutive positive integers and are used as a logical clock\,\cite{Lamport:1978p21405}, allowing servers to detect obsolete information.
Each server stores its current term number, which increases monotonically over time.
This number is exchanged by communicating servers, allowing servers to update to the more recent value.
If a server receives a request with an older term number, it is rejected, and its own current term number is sent back to the contacting server.
If a candidate or a leader receives such a response, it immediately reverts to the follower state.

Raft uses the leader election as the first of two phases of consensus, using a heartbeat mechanism to trigger it.
All servers start up as followers, and wait to be contacted by the leader or a candidate for a period of time equivalent to the election timeout.
When this election timeout elapses, because it has not received any valid invocations from a leader or a candidate, the server, assuming there is no leader on the cluster, becomes a candidate. It will then increase its term value, reset the election timeout, by assigning a randomly selected value to help prevent split votes, and issue RequestVote RPCs in parallel to all known servers, starting a new leader election.
This candidate will be elected as the new leader after receiving a majority of votes from the servers comprising the cluster.

A leader uses this very same election timeout to trigger periodic heartbeats, that correspond to issuing AppendEntries RPCs that contain no log entries to all of its followers, in order to keep its authority.
If a leader fails or becomes disconnected a new one is elected.

The leader will then assume full responsibility for managing the replicated log.
Hence, it accepts client requests, that contain a command to be executed by the replicated state machines, which is converted into an entry added to its log.
Afterwards, the leader issues AppendEntries RPCs in parallel to its known followers, in order to replicate the entry.
When it has been safely replicated, i.e. received a number of responses that is equal to the majority of the elements of the cluster, it applies the entry to its state machine and returns the result of that execution to the client.
The leader will inform the replicas to commit that entry to their state machines in subsequent AppendEntries invocations.

The Raft algorithm ensures the replicated state machine safety property, which states that if any server has applied a particular log entry to its state machine, no other server may apply a different command for the same log index.

The failure of a follower or a candidate is easily dealt by the Raft protocol, as RequestVote or AppendEntries RPCs sent to it will fail. But when the server restarts as a follower, the RPCs will be delivered and processed correctly.

The normal interaction of a client with a Raft cluster is as follows.
The client randomly selects a server and sends it a request.
If that server is the leader it will process the request and return the response to the client.
If the server is not the leader, it sends the address of the most recent known leader to the client, which can use it to contact the leader directly.
In the case the leader crashes, the client's request will time out and it will need to select another server to interact with.

The authors of the Raft consider it an easier protocol to understand than Paxos\,\cite{279229}, fact supported by a user study where the majority of the inquired subjects considered Raft easier than Paxos to implement and to explain\,\cite{Ongaro:2014p20967}.
Another drawback of Paxos is the lack of a reference algorithm for multi-Paxos, as most descriptions fall on single-decree Paxos or leave too many details to the implementer\,\cite{Ongaro:2014p20967}.
Compared to ZooKeeper, which is also leader-based, Raft is also a simpler protocol as it requires the implementation of fewer distinct operations and it also minimizes the functionality in non-leaders.
For instance, the log entries flow in a single direction, from the Raft leader to its replicas, whereas in ZooKeeper, entries flow both to and from the leader.

\section{Implementation}
\label{sec:proposal}
Our proposal is to apply the Raft protocol on the DPWS environment, and therefore, we have implemented a Raft Service on top of the Web Services for Devices (WS4D) Java Multi Edition DPWS Stack (JMEDS), which will be described throughout this section.
Our framework considers two different entities: Servers, which host an instance of the Raft Service, and Clients, which invoke those instances.

\subsection{Server}
A Server is a device with the type ``Raft\_Device'', using the DPWS terminology, and so, the terms Raft Device and Server will be used interchangeably throughout this section.
The Server class contains five different entities, Log, Raft Service, ServerClient, TimeoutTask and the current state task.
It also stores Raft specific parameters, such as the current term (\textit{currentTerm}), the Server it has voted for (\textit{votedFor}), the Server that is the current leader (\textit{currentLeader}), the next index (\textit{nextIndex}) and the match index (\textit{matchIndex}) for each replica or follower.

Essential for the replicated state machine approach, the Log keeps a list with all the entries, each represented by a LogEntry object, resultant of the commands inserted by clients, as well as the StateMachine and some variables, like the \textit{commitIndex} and the \textit{lastApplied}, which correspond to the highest log entry known to be committed or applied to the StateMachine, respectively.
Each Server periodically runs a task to compare the values of \textit{commitIndex} and \textit{lastApplied}. If \textit{commitIndex} is bigger, \textit{lastApplied} is incremented, and Log[\textit{lastApplied}], i.e. the LogEntry with an index equal to \textit{lastApplied}, is committed, by applying it to the StateMachine.
The StateMachine provides operations for its initialization, termination, and for the insertion of a LogEntry returning a boolean value to convey the success of the insertion.\\
Before describing the operations provided by the Raft Service we will better describe the structure of the LogEntry.
Each new LogEntry is created by the leader and it stores the following parameters:
\begin{description}
\item[\textit{index}] Unique index assigned by the leader to be the size of Log (henceforth identified by \textit{lastLogIndex}) plus 1, which corresponds to the successor of the index of the last log entry, as the index of the first entry is 1.
\item[\textit{term}] \textit{currentTerm} of the leader.
\item[\textit{uid}] Unique identifier extracted from the client's request.
\item[\textit{command}] Operation to be executed on the StateMachine.
\item[\textit{parameters}] Parameters for the operation defined by the command argument.
\item[\textit{result}] Result of the execution of the LogEntry's \textit{command}.
\item[\textit{success}] Success indicator of the execution of the LogEntry's \textit{command}.
\end{description}

Besides these parameters, each LogEntry object also stores the number of follower responses needed to achieve the majority, according to the current size of the Raft cluster, and the respective lock, which will unlock when the majority is reached.
These responses are added to a LogEntry upon the reception of successful follower responses to the invocation of the AppendEntries operation containing LogEntry.
After unlocking, the answer is sent back to the client that issued the request originally.

The Raft Service provides 3 different operations, which match very closely Raft's RPCs, and are available on every instance:
\begin{description}
\item[InsertCommand] Invoked by clients to insert new commands in the cluster's replicated log.
A request to this operation takes the following arguments:
\begin{description}
\item[uid] Unique identifier generated by the invoking client to identify its request.
\item[command] Operation to be executed on the replicated state machine.
\item[parameters] Parameters for the operation defined in \textbf{command}.
\end{description}
The response of this operation comprises the arguments:
\begin{description}
\item[success] Indicator of the success of the command execution by the leader, which is always false if the Server is not the current leader.
\item[result] Result of applying the command if the Server is the current leader.
\item[leaderAddress] Conveys the contacted Server's \textit{currentLeader} if it is a follower.
\end{description}
\end{description}
Upon receiving a request, the leader verifies if there is no LogEntry on the Log corresponding to \textbf{uid}, on which case the LogEntry's \textit{result} will be sent right back to the client indicating a successful execution.
Otherwise, the leader's Log creates the new LogEntry and sets the number of responses necessary to unlock its \textit{result}, and the TimeoutTask is notified in order to trigger the invocation of the AppendEntries operation in order to propagate LogEntry to the followers.
When the LogEntry is unlocked, the response is sent back to the client containing the \textbf{success} and the \textbf{result} of creating and committing LogEntry.
If the Server is not the current leader of the Raft cluster, it responds right away to the client with \textbf{success} equal to false, and \textbf{leaderAddress} equal to \textit{currentLeader}.

\begin{description}
\item [AppendEntries] Invoked by the leader as an heartbeat, when there are no new entries, or to replicate new log entries on its followers.
A request to this operation takes the following arguments:
\begin{description}
\item [term] Leader's \textit{currentTerm}.
\item [leaderId] The address of the Raft Service of the leader (\textit{currentLeader}).
\item [prevLogIndex] The index of the log entry that precedes the new ones.
\item [prevLogTerm] The term of the log entry identified by \textbf{prevLogIndex}.
\item [entries] The list of log entries to store, which will be empty in case the message is an heartbeat. Each entry is defined through its \textbf{index}, \textbf{term}, \textbf{uid}, \textbf{command} and \textbf{parameters} arguments.
\item [leaderCommit] Leader's \textit{commitIndex}.
\end{description}
The response of this operation comprises the arguments:
\begin{description}
\item [term] \textit{currentTerm} of the targeted Server, for the leader to update itself.
\item [success] Boolean value indicating if the Log of the targeted Server contains the entry matching the values of \textbf{prevLogIndex} and \textbf{prevLogTerm}.
\end{description}
\end{description}

Independently of the success of this operation, the targeted Server returns its \textit{currentTerm} on the reply.
It starts by extracting the received \textbf{term}, and comparing it to \textit{currentTerm}.
If it is smaller, the Server will immediately send back a response with false on \textbf{success}.
Otherwise, it checks if its Log contains a LogEntry with an index equal to \textbf{prevLogIndex}, and the \textit{term} equal to \textbf{prevLogTerm}.
If such a LogEntry does not exist, the \textbf{success} argument will be false and no further processing of the request takes place.
However, if it does, the \textbf{success} argument of the response is set to true and the address of the leader, \textbf{leaderId}, will be copied to \textit{currentLeader} if it is different from the previous value of \textit{currentLeader}.
The Server will notify its current state task by invoking its heartbeat method and will extract the entries from the request, reconstructing each LogEntry from its \textbf{index}, \textbf{term}, \textbf{uid}, \textbf{command} and \textbf{parameters} arguments, which will then be sent to the Log so they are appended to the existing entries.
During this process, if there is any LogEntry in the Log that conflicts with a new one, by having the same \textit{index} but different \textit{terms}, it will be deleted as well as of all the subsequent existing ones.
Any new LogEntry not in the Log is inserted in the corresponding \textit{index}.
Before returning its response to the leader, and if the request processing has been successful so far, the Server compares the value of the \textbf{leaderCommit} argument to its \textit{commitIndex}.
If its \textit{commitIndex} is inferior, it takes the minimum value between \textbf{leaderCommit} and its \textit{lastLogIndex}.
Finally, the response is sent back to the leader conveying the Server's \textit{currentTerm} on \textbf{term}, and the \textbf{success} of the request.

\begin{description}
\item [RequestVote] Invoked by candidates to gather votes from other Servers on the cluster.
A request to this operation takes the following arguments:
\begin{description}
\item [term] Candidate's \textit{currentTerm}.
\item [candidateId] The address of the Raft Service of the candidate.
\item [lastLogIndex] The index of the last log entry of the candidate or \textit{lastLogIndex}.
\item [lastLogTerm] The term of the last log entry of the candidate.
\end{description}
The response of this operation comprises the arguments:
\begin{description}
\item [term] \textit{currentTerm} of the targeted Server.
\item [voteGranted] Boolean value indicating if the Server has voted, or not, for this candidate to become the new leader.
\end{description}
\end{description}
The targeted Server compares \textbf{term} with its \textit{currentTerm}.
If it is smaller, the Server will immediately send back a response with \textit{currentTerm} on the \textbf{term} argument, and the false value on the \textbf{voteGranted} argument.
Otherwise, it compares \textbf{candidateId} with the Server's \textit{votedFor}.
If they are equal or \textit{votedFor} is null, the Server extracts and analyzes the remaining request arguments.
It verifies if the candidate's Log is as up-to-date or more advanced than its own Log, by checking if the values of \textbf{lastLogIndex} and \textbf{lastLogTerm} are greater than or equal to its corresponding parameters, \textit{lastLogIndex} and the term of Log[\textit{lastLogIndex}], respectively.
If this conditions are confirmed, the Server sets its \textit{votedFor} variable to \textbf{candidateId}, granting it the vote.
After this verifications, the reply to the candidate conveys the Server's \textit{currentTerm} value on \textbf{term}, and \textbf{voteGranted} will indicate if the Server has granted or not its vote to the candidate.

In the AppendEntries and RequestVote operations, the involved Servers always compare the value of the received \textbf{term} argument with their own \textit{currentTerm}.
If the received value is higher, the Server sets its \textit{currentTerm} to \textbf{term} and converts to the follower state, if it wasn't in that state previously.

After describing the Raft Service and its operations, the remaining components of a Server are described, starting by its associated ServerClient which can be used to detect other Servers, by listening to WS-Discovery multicast messages or issuing Probe messages to find other Raft Devices.
All the detected Raft Devices will be queried to retrieve the AppendEntries and the RequestVote operation stubs, which are stored alongside with the matching Raft Service address, in case the current Server receives an invocation to its InsertCommand operation while it isn't a leader and it is necessary to return the leader's Raft Service address to the client.
This information is stored and indexed by the endpoint reference of the device.
The detection of multicast Bye messages sent by a known Raft Device, makes the ServerClient remove the corresponding information on that device.
Besides this function of maintaining the information on the cluster's elements, the ServerClient is the Server's component used to invoke operations on other Servers, such as RequestVote, when it becomes a candidate and starts a new election, or AppendEntries, when it is a leader and must signal its liveness, using heartbeats, or must replicate log entries on its followers.
Such invocations are performed in parallel, by using a thread for invoking an operation on each known Server, following the guidelines of the Raft algorithm.

Each Server has a TimeoutTask thread that runs throughout the entire life of the Server, using a loop that is stopped when the Server shuts down.
On each cycle, the TimeoutTask starts by checking if the state of the Server should be altered and performs the transition, if necessary.
Afterwards, this task waits for a period of time corresponding to the configured election timeout value.
When this time interval elapses, the TimeoutTask invokes the timeout method of the current state's task object.
The timer will be interrupted, avoiding the mentioned invocation, by received invocations to the AppendEntries operation in both follower and candidate states, or to the RequestVote operation, in case the follower grants its vote, or when a majority of votes is received by the current Server while being a candidate.

According to the Raft protocol, a server can be in one of three different states, follower, candidate or leader, and it always starts its lifecycle as a follower.
These states are represented by the FollowerTask, CandidateTask and LeaderTask classes which extend the ServerTask abstract class, making them share a basic interface with operations that are common to all the states, such as the reception of an invocation to the AppendEntries operation, the elapsing of the election timeout, or the termination of the current state.
The signaling of the reception of an invocation to the AppendEntries is made through the heartbeat method.
In the case of the FollowerTask, the heartbeat method notifies the TimeoutTask to interrupt the currently waiting timer and skip to the next cycle, hence restarting the timer.
The heartbeat methods of both the CandidateTask and the LeaderTask behave similarly, merely setting the Server's next state as follower, before notifying the TimeoutTask.
The invocation of the timeout method on the LeaderTask will cause the invocation of the AppendEntries operation on the known replicas.
These invocations will convey any new entries received from the last invocation, or none if the leader's log hasn't been modified.
The timeout method of the FollowerTask informs the TimeoutTask to make the Server transition to the candidate state.
On the CandidateTask, the timeout method leads to the execution of a new election.

\subsection{Client}
The normal execution of a simple client for the Raft Service is to detect Raft Devices, whether by registering to listen to multicast WS-Discovery Hello messages from such devices, or to actively search for devices with such a type by issuing a multicast Probe message.
If any Raft Devices are in the same network, they will respond to the client with a ProbeMatch message.

All the detected Raft Devices, whichever the used discovery mechanism, will be queried to retrieve the InsertCommand operation stub, which is stored with the matching device endpoint reference.
The first detected Raft Device will be considered as the leader by the client, which will then become the target for its invocations of the InsertCommand operation.

Let's look into such an invocation in detail.
A client prepares the invocation of the InsertCommand operation through the previously retrieved leader stub.
It requests the creation of a Universally Unique Identifier (UUID) to the IDGenerator class provided by the WS4D JMEDS.
This UUID, as well as the desired command and parameters are inserted in the request message that is then sent to the leader's InsertCommand operation.
If the client's selected target is the current leader of the Raft cluster, the response will convey whether the creation and application of the corresponding LogEntry to the leader's StateMachine was successful and its result.
Otherwise, i.e. if the targeted InsertCommand operation belongs to a Raft Device or Server that is currently a follower, the response will convey the false value as well as the address of the Raft Service instance of the current leader.
In this case, the client can then extract this address to reissue the invocation to the correct cluster leader, in order for it to become effective, as well as to send it further invocations.

\section{Performance Evaluation}
\label{sec:results}
To evaluate the performance of our implementation of the Raft protocol using version 2 beta 10 of Java Multi Edition DPWS Stack (JMEDS), part of the Web Services for Devices (WS4D) project\,\cite{ws4d_url}, it was decided to compare it with ZooKeeper in similar scenarios with a single server, and three or five servers, to process the requests of 1, 5, 10 or 25 clients.
Five runs were executed for each scenario and the average of their results is presented.

We have leveraged WS-Discovery's inclusion on DPWS by having a test manager listen to multicast announcements of clients and servers entering the network, to determine if all the intervening components were up and running in order to start the execution of test.
Hence, the execution procedure of each Raft test comprised the following steps:
\begin{enumerate}
 \item The manager and the clients are started on the same machine.
 \item Each server is then started on its own machine.
 \item The manager awaits the detection of all the expected clients and servers, for the current test.
 The manager will then select the leader server and contact each client to inform about it and also on the number of iterations to execute.
 The manager will also inform each server about its state, if it is a leader or a follower, its neighboring servers, and the value of the election timeout, which is randomly selected from 150 to 300 milliseconds.
 After conveying all the relevant parameters, the manager requests all servers to start running and waits for twice the value of the election timeout, before subscribing to the end of workload event on all the clients.
 Afterwards, starts the workload on all clients.
 \item The clients start executing the specified number of iterations where they invoke the InsertCommand operation on the leader server.
 \item The clients terminate and notify the manager, which waits until all the clients have notified it.
 \item The manager informs all the clients and servers on the name of the file they should write their run statistics to.
 \item All clients and servers terminate after writing the statistics.
\end{enumerate}

The effects of failing servers were also evaluated, until the maximum number of tolerated failures, as the service should become unavailable in order to guarantee its correctness.
For this purpose, the manager was configured to cause a failure at 500 milliseconds after the start of the workload, which could be a follower or the leader in the scenarios with 3 servers.
These two scenarios will be hence forth identified by 1 Follower (1F) and 1 Leader (1L), respectively.
In the failure scenarios with 5 servers, a follower failure is introduced at the same instant of time, i.e. 500 milliseconds, followed by another failure, after another 500 milliseconds, which can be another follower or the leader.
These will be hence forth identified by 2 Followers (2F) and 1 Follower 1 Leader (1F1L), respectively.
The baseline scenarios, when compared with the failing ones, will be identified by 0 Errors (0E) to better distinguish them.

In order to better compare the behavior of a client reconnecting to the cluster after losing its current connection, we have mimicked the behavior of the \textit{ClientCnxn} class as provided in ZooKeeper.
When a loss of connection is detected, it waits for an interval of time, whose dimension is randomly generated until a maximum of 1 second, before attempting to connect to another server.

We compared version 3.4.5 of Apache ZooKeeper with our implementation of Raft. The configuration parameters used for all the settings were the following: a tickTime of 2000 milliseconds, which corresponds to ZooKeeper's basic time unit or heartbeat interval; an initLimit of 5, which means ZooKeeper servers have 5 ticks to connect to a leader; and a syncLimit of 2 ticks, which is the maximum delay of the state of a ZooKeeper server compared to the quorum's leader.
The procedure for each ZooKeeper test was exactly the same as described for the Raft tests, with the exception of the interactions between the manager and the servers, since we did not want to modify the code of the ZooKeeper servers.
Hence, we initialized each ZooKeeper server using the supplied \textit{zkServer} bash script, and only started the test manager and the clients after waiting for the time corresponding to the initLimit parameter.

%
%
%

\subsection{Experimental settings}
The experimental evaluation of our implementation of the Raft protocol compared to the Apache ZooKeeper was performed on six identical hosts connected to the same LAN, with the following configuration: 64-bit Ubuntu 12.04.4 Lin{-}ux, Intel\textsuperscript{R} Core\textsuperscript{TM} i3-2100, 3.10GHz, 8GB RAM, 64-bit Java\textsuperscript{TM} SE 1.6.0\_27.
One machine was exclusively used to run the manager and all the clients, whereas each of the remaining machines was used to run a single Raft or Zookeeper server, in sets of 1, 3 or 5 servers according to the tests' settings.

Each client executed 120 iterations without any interval, where each iteration consists on invoking the insertion of a new command on the leader server in the case of Raft, or a randomly selected server in the case of ZooKeeper, as it is the default behavior of the ZooKeeper client.
Each command contains a unique identifier, defined by each invoking client, as well as the actual command and the corresponding parameters.
The leader, after receiving such a request, creates the corresponding entry on its log and invokes the AppendEntries operation of the Raft service on its known replicas, to propagate the new entry.
The leader will only respond to the client when the majority of its replicas has replied successfully to this invocation.
For the execution of the Raft Service tests, the selected state machine was the one built using Berkeley DB.

In the case of ZooKeeper, the insertion of a command corresponds to the creation of a file with the unique identifier as its name, with the command and the parameters as its contents.
The ZooKeeper server replies with the full file path to the client.
Before each run, all the created files were deleted, in order to start with an empty file-system.
For latency measurements, the first and the last 10 iterations were discarded in order to minimize the effect of Java JIT compilation, although it also masks the delay of TCP connection establishment, whereas throughput takes into account all the 120 iterations, where clients issued requests, and the time it took servers to process all of them and to reply back.

\subsection{Results and Discussion}

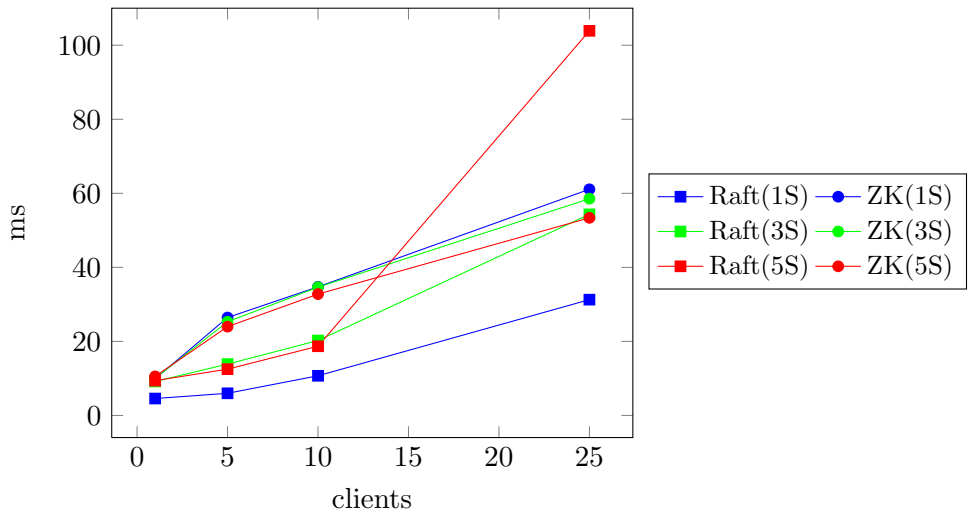
\begin{figure}[htbp]
  \centering
\begin{tikzpicture}
 \begin{axis}[xlabel=clients,ylabel=ms,ymax=110,legend columns=2,
 legend style={
 	at={(1.03,0.48)},
	anchor=west,font=\small}]
  \addplot[color=blue, mark=square*] table[header=false,x index=0,y index=1]{stats/raft_1s.csv};
  \addplot[color=blue, mark=*] table[header=false,x index=0,y index=1]{stats/zoo_1s.csv};
  \addplot[color=green, mark=square*] table[header=false,x index=0,y index=1]{stats/raft_3s.csv};
  \addplot[color=green, mark=*] table[header=false,x index=0,y index=1]{stats/zoo_3s.csv};
  \addplot[color=red, mark=square*] table[header=false,x index=0,y index=1]{stats/raft_5s.csv};
  \addplot[color=red, mark=*] table[header=false,x index=0,y index=1]{stats/zoo_5s.csv};
  \legend{Raft(1S),ZK(1S), Raft(3S),ZK(3S), Raft(5S),ZK(5S)}
 \end{axis}
\end{tikzpicture}
\caption{Raft vs. ZooKeeper (Latency).}
\label{fig:lats}
\end{figure}

\begin{figure}[htbp]
  \centering
\begin{tikzpicture}
 \begin{axis}[xlabel=clients,ylabel=Ops/s,ymax=800, legend columns=2,
 legend style={
 	at={(1.03,0.48)},
	anchor=west,font=\small}]
  \addplot[color=blue, mark=square*] table[header=false,x index=0,y index=2]{stats/raft_1s.csv};
  \addplot[color=blue, mark=*] table[header=false,x index=0,y index=2]{stats/zoo_1s.csv};
  \addplot[color=green, mark=square*] table[header=false,x index=0,y index=2]{stats/raft_3s.csv};
  \addplot[color=green, mark=*] table[header=false,x index=0,y index=2]{stats/zoo_3s.csv};
  \addplot[color=red, mark=square*] table[header=false,x index=0,y index=2]{stats/raft_5s.csv};
  \addplot[color=red, mark=*] table[header=false,x index=0,y index=2]{stats/zoo_5s.csv};
  \legend{Raft(1S),ZK(1S), Raft(3S),ZK(3S), Raft(5S),ZK(5S)}
 \end{axis}
\end{tikzpicture}
\caption{Raft vs. ZooKeeper (Throughput).}
\label{fig:throughput}
\end{figure}
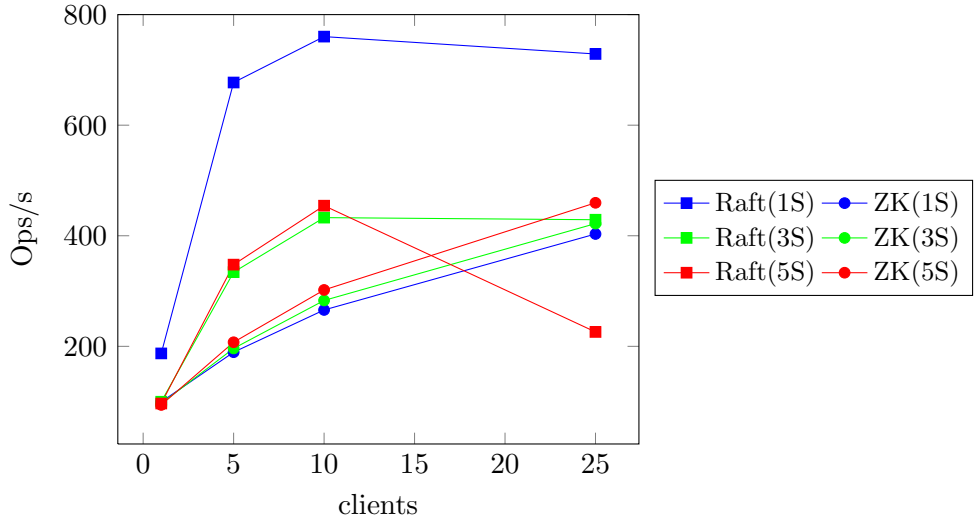

In Figure\,\ref{fig:lats}, the latency of all the scenarios grows linearly with the number of clients, with the exception of Raft with 5 servers, where the latency seems to increase exponentially from 10 to 25 clients, possibly, showing signs of saturation of the cluster's resources.
The latency of the baseline scenario, i.e. Raft with a single server which corresponds to the non-replicated service, is, as expected, noticeably inferior to the remaining scenarios, independently of the number of clients.
It is important to notice that the latency of the various ZooKeeper scenarios, decreases as the number of servers increases.
Another important fact is that the latency of the Raft scenarios with both 3 and 5 servers is very similar and inferior to the corresponding ZooKeeper scenario, with the exception of the case mentioned before, i.e., Raft with 5 servers and 25 clients, which reaches an average of around 103 milliseconds.
Figure\,\ref{fig:throughput} shows the throughput in operations per second that the servers can fulfill in the various scenarios, which is closely related with the latency values depicted in Figure\,\ref{fig:lats}.
Regarding the ZooKeeper scenarios, one can see that the throughput increases with the number of servers, which supports the known parallelism and high-availability capabilities of ZooKeeper, as clients can be distributed throughout the various servers of the cluster.
On the other hand, as the Raft protocol only allows the leader to satisfy client requests, which consist in the insertion of commands in these tests, the scalability of this protocol suffers from this limitation, as the leader can easily become overloaded since it processes all the updates to the state machine, by propagating them to the followers, and needs to respond to the connected clients.

\begin{figure}[htbp]
  \centering
\begin{tikzpicture}
 \begin{axis}[xlabel=clients,ylabel=ms,ymax=82, legend columns=2,
 legend style={
 	at={(1.03,0.48)},
	anchor=west,font=\small}]
  \addplot[color=blue, mark=square*] table[header=false,x index=0,y index=1]{stats/raft_3s.csv};
  \addplot[color=blue, mark=*] table[header=false,x index=0,y index=1]{stats/zoo_3s.csv};
  \addplot[color=green, mark=square*] table[header=false,x index=0,y index=1]{stats/raft_3s_1replica.csv};
  \addplot[color=green, mark=*] table[header=false,x index=0,y index=1]{stats/zoo_3s_1replica.csv};
  \addplot[color=red, mark=square*] table[header=false,x index=0,y index=1]{stats/raft_3s_1leader.csv};
  \addplot[color=red, mark=*] table[header=false,x index=0,y index=1]{stats/zoo_3s_1leader.csv};
  \legend{Raft(0E),ZK(0E),Raft(1F),ZK(1F),Raft(1L),ZK(1L)}
 \end{axis}
\end{tikzpicture}
\caption{Raft vs. ZooKeeper with 3 servers with a failed server at 500 ms (Latency).}
\label{fig:lats_3s}
\end{figure}
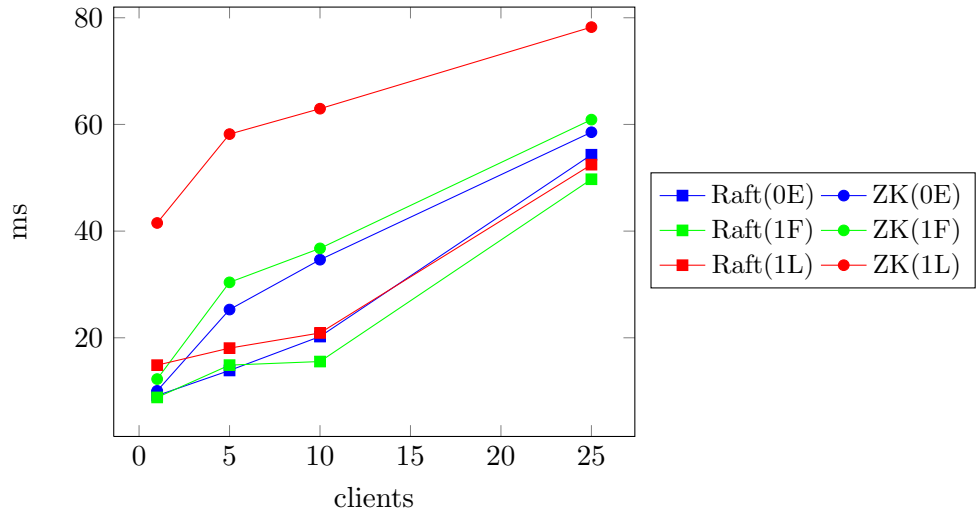

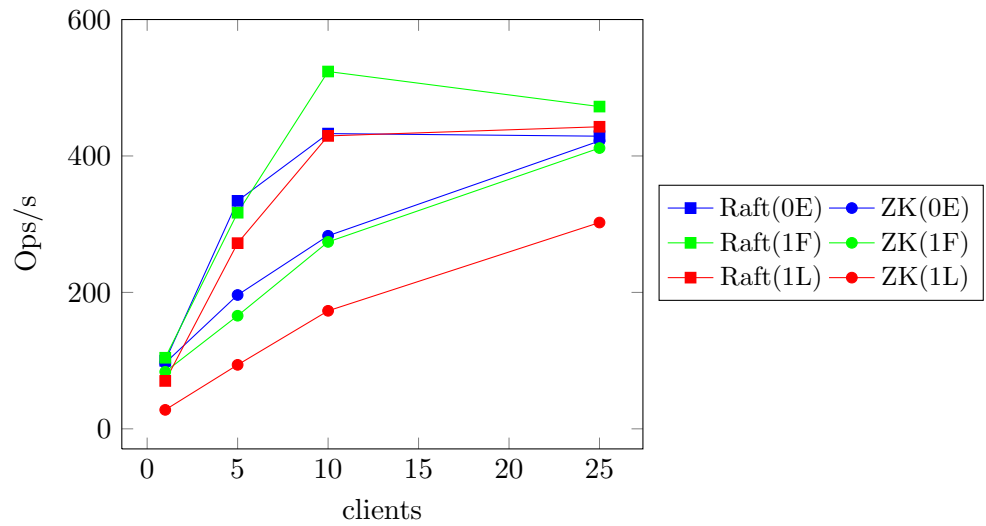
\begin{figure}[htbp]
  \centering
\begin{tikzpicture}
\begin{axis}[xlabel=clients,ylabel=Ops/s,ymax=600, legend columns=2,
 legend style={
 	at={(1.03,0.48)},
	anchor=west,font=\small}]
  \addplot[color=blue, mark=square*] table[header=false,x index=0,y index=2]{stats/raft_3s.csv};
  \addplot[color=blue, mark=*] table[header=false,x index=0,y index=2]{stats/zoo_3s.csv};
  \addplot[color=green, mark=square*] table[header=false,x index=0,y index=2]{stats/raft_3s_1replica.csv};
  \addplot[color=green, mark=*] table[header=false,x index=0,y index=2]{stats/zoo_3s_1replica.csv};
  \addplot[color=red, mark=square*] table[header=false,x index=0,y index=2]{stats/raft_3s_1leader.csv};
  \addplot[color=red, mark=*] table[header=false,x index=0,y index=2]{stats/zoo_3s_1leader.csv};
  \legend{Raft(0E),ZK(0E),Raft(1F),ZK(1F),Raft(1L),ZK(1L)}
 \end{axis}
\end{tikzpicture}
\caption{Raft vs. ZooKeeper with 3 servers with a failed server at 500 ms (Throughput).}
\label{fig:throughput_3s}
\end{figure}

The effects of a failed server, at 500 milliseconds, in a cluster with 3 servers can be observed in Figure\,\ref{fig:lats_3s} and Figure\,\ref{fig:throughput_3s}.
The average latency of the ZooKeeper scenarios is always higher than in the corresponding Raft scenario, and, in all these scenarios, it increases linearly with the number of clients.
In terms of throughput, the ZooKeeper scenarios seem to increase linearly opposed to the Raft ones, where a peak is reached in the runs with 10 clients, with the throughput decreasing slightly or stabilizing afterwards, which seems to show that the plateau of the processing capabilities of the leader has been reached.

The failure of a ZooKeeper server, alway increases the latency and, consequently, reduces throughput.
Whereas the failure of a follower only deteriorates the performance slightly, from 2 to 5 milliseconds in latency and from 10 to 30 operations per second in throughput, the failure of the leader introduces a penalty of 20 to 30 milliseconds in latency and 70 to 120 operations per second in throughput.

The effects of a failed Raft server vary.
A failed follower has better performance than the baseline, from decreasing the latency between 1 and 5 milliseconds, and increasing the throughput between 5 and 90 operations per second, except for 5 clients, where it is slightly worse.
This can be explained by the smaller number of messages a leader needs to send, as it only needs to contact a single follower, instead of two as in the baseline.
The scenario with the failed leader implies that all clients connect to the newly elected leader, to fulfill its requests, having a worse performance, as occurs more distinctively for 1 and 5 clients, and in a smaller degree for 10 clients.
For 25 clients, its performance is better than the baseline which could be explained with the same phenomenon caused by a failed follower, i.e., as the leader is killed, a follower will eventually be elected as the new leader, which will need to contact the single follower remaining in the cluster, hence issuing a smaller number of messages.
This will certainly counterbalance the penalty introduced by the death of the leader, which causes clients to connect to the new leader.

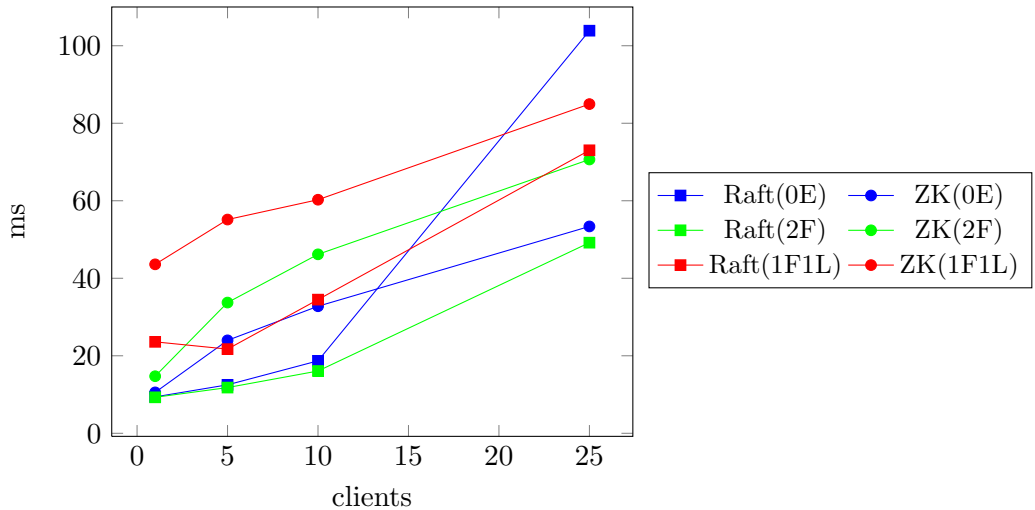
\begin{figure}[htbp]
  \centering
\begin{tikzpicture}
 \begin{axis}[xlabel=clients,ylabel=ms,ymax=110, legend columns=2,
 legend style={
 	at={(1.03,0.48)},
	anchor=west,font=\small}]
  \addplot[color=blue, mark=square*] table[header=false,x index=0,y index=1]{stats/raft_5s.csv};
  \addplot[color=blue, mark=*] table[header=false,x index=0,y index=1]{stats/zoo_5s.csv};
  \addplot[color=green, mark=square*] table[header=false,x index=0,y index=1]{stats/raft_5s_2replicas.csv};
  \addplot[color=green, mark=*] table[header=false,x index=0,y index=1]{stats/zoo_5s_2replicas.csv};
  \addplot[color=red, mark=square*] table[header=false,x index=0,y index=1]{stats/raft_5s_1replica1leader.csv};
  \addplot[color=red, mark=*] table[header=false,x index=0,y index=1]{stats/zoo_5s_1replica1leader.csv};
  \legend{Raft(0E),ZK(0E),Raft(2F),ZK(2F),Raft(1F1L),ZK(1F1L)}
 \end{axis}
\end{tikzpicture}
\caption{Raft vs. ZooKeeper with 5 servers with two failed servers at 500 and 1000 ms (Latency).}
\label{fig:lats_5s}
\end{figure}

\begin{figure}[htbp]
  \centering
\begin{tikzpicture}
 \begin{axis}[xlabel=clients,ylabel=Ops/s,ymax=600, legend columns=2,
legend style={
 	at={(1.03,0.48)},
	anchor=west,font=\small}]
  \addplot[color=blue, mark=square*] table[header=false,x index=0,y index=2]{stats/raft_5s.csv};
  \addplot[color=blue, mark=*] table[header=false,x index=0,y index=2]{stats/zoo_5s.csv};
  \addplot[color=green, mark=square*] table[header=false,x index=0,y index=2]{stats/raft_5s_2replicas.csv};
  \addplot[color=green, mark=*] table[header=false,x index=0,y index=2]{stats/zoo_5s_2replicas.csv};
  \addplot[color=red, mark=square*] table[header=false,x index=0,y index=2]{stats/raft_5s_1replica1leader.csv};
  \addplot[color=red, mark=*] table[header=false,x index=0,y index=2]{stats/zoo_5s_1replica1leader.csv};
  \legend{Raft(0E),ZK(0E),Raft(2F),ZK(2F),Raft(1F1L),ZK(1F1L)}
 \end{axis}
\end{tikzpicture}
\caption{Raft vs. ZooKeeper with 5 servers with two failed servers at 500 and 1000 ms (Throughput).}
\label{fig:throughput_5s}
\end{figure}
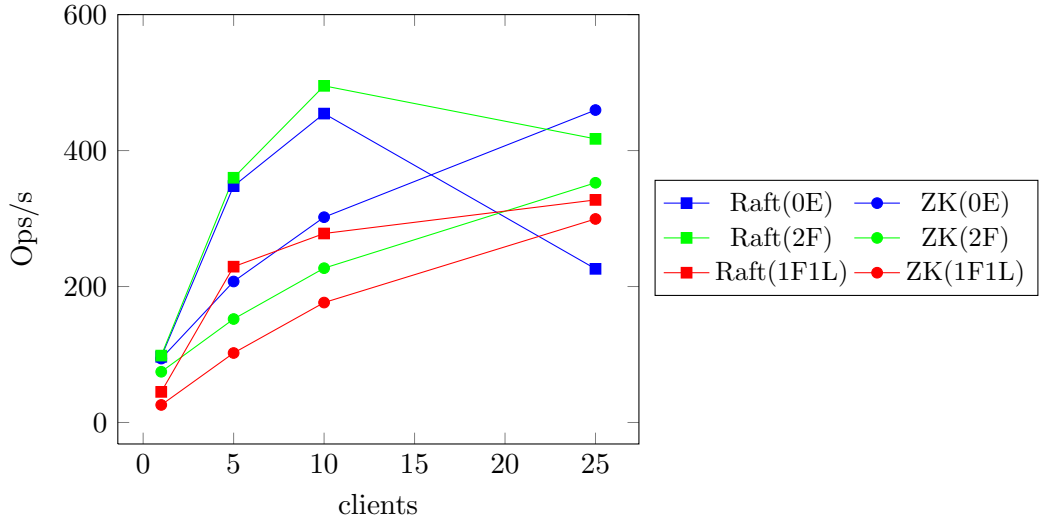

Figure\,\ref{fig:lats_5s} and Figure\,\ref{fig:throughput_5s} portray the influence of 2 failed servers, a follower at 500 milliseconds, and another follower or the leader at 1000 milliseconds, in a cluster with 5 servers, reaching the maximum number of tolerated failures.
As occurred with the cluster of 3 servers, the average latency of the ZooKeeper scenarios is always higher than the corresponding Raft scenario, which increases linearly with the number of clients in all the scenarios.
The only exception, as mentioned previously in the comments to Figure\,\ref{fig:lats} and Figure\,\ref{fig:throughput}, was the baseline scenario, Raft(0E), with 25 clients.

Failed ZooKeeper servers always introduce a performance penalty, most probably due to the need of clients that were connected to the failing server, to reconnect to a different one in order to invoke their requests, in both failure scenarios.
The latency increases between 4 and 17 milliseconds, and the throughput decreases between 9 and 31 operations per second, in the ZK(2F) scenario.
The ZK(1F1L) scenario introduces an aggravated penalty, as the latencies are around 30 milliseconds higher and the throughput decreases between 20 and 130 operations per second compared to the baseline, the ZK(0E) scenario.

The effects of failed servers in a Raft cluster with 5 servers vary in a similar way to what was observed for 3 servers.
The failure of 2 followers improves the performance, by reducing ever so slightly the latency, and increasing the throughput between 2 to 40 operations per second, with the exception of 25 clients, where the cluster is clearly saturated in the baseline, which worsens its overall performance dramatically.
This exact same setting for the baseline is the only one where the Raft(1F1L) scenario has a better performance.
On the remaining settings, Raft(1F1L) introduces an overhead varying from 9 to 16 milliseconds in terms of latency and a reduction of around 50 to 180 operations per second in throughput.

To sum it all up, the performance of Raft is always better than that of ZooKeeper with similar settings, excepting the baseline Raft scenario with 5 servers, which shows signs of resources saturation when serving 25 clients.
However, Figure\,\ref{fig:throughput} shows a trend that the throughput for ZooKeeper is still increasing, which could continue past the 25 clients, whereas the throughput of Raft clusters seems to have stagnated around 430 operations per second.
Albeit allowing clients to connect to followers, hence sharing the load of processing their requests, by propagating them to the quorum and answering back to clients, ZooKeeper suffers more from failed followers, as the clients connected to a failed follower will need to connect to another server in the quorum to invoke subsequent requests.
On the contrary, Raft's performance increases in a similar failure scenario, as the leader needs to contact a smaller number of followers and the clients will only need to reconnect when the leader fails.
The failure of the leader causes an additional aggravation of the performance, as in Raft it leads to clients reconnecting to the new leader, and in ZooKeeper, the service becomes unavailable until the new leader has been elected, and only then the clients will be able to reconnect to a server in the cluster.

\section{Related Work}
\label{sec:related}

WS-Replication\,\cite{1135831} offers transparent active replication of Web Services, and relies on WS-Multicast to achieve multicast communication with the replicas and to perform node failure detection using a SOAP-based ping mechanism.
WS-Multicast can also be used as a standalone component in order to provide reliable multicast in a Web Service environment\,\cite{1135831}.
The proposed framework allows the deployment of a Web Service in a set of sites to increase its availability, and transparently forwards a normal web service invocation to its replicas using multicast.
The reply to this invocation is sent back to the client when the service receives the configured number of responses, which could be one, a majority or all. 

Primary-backup, or passive, replication has also been applied to Web Services\,\cite{4279648, jayasinghe2005faws,Liang:2003p2721}, but, due to its modular design, the framework presented in\,\cite{4279648} must be stressed, as it is easily extendable to use active or coordinator-cohort replication.

Byzantine Fault Tolerance (BFT)\,\cite{Lamport:1982p21428} is a replication technique designed to protect against arbitrary problems like crash faults, software bugs or security violations and requires a higher degree of replication than crash faults tolerant techniques.
Thema\,\cite{Merideth:2005p2739} provides a structured way to build Byzantine-fault-tolerant and survivable Web Services that are externally visible and accessible as standard Web Services, being able to attend non-replicated clients and to access non-replicated Web Services.
Thema incorporates the Castro-Liskov Practival Byzantine Fault Tolerance (CLBFT)\,\cite{castro1999practical} protocol, in order to achieve a reliable and secure transport layer without any synchrony assumptions for safety.
This middleware system provides SOAP and WSDL support for BFT, as well as adding multi-tier support to BFT, while working in a mixed-fault model. 

BFT-WS\,\cite{Zhao:2007p2923} is a Byzantine fault tolerance middleware framework for Web Services, and, like Thema, it is based on CLBFT.
It builds upon WS-ReliableMessaging to achieve reliable control communication.
As Thema uses a wrapper to interface with a BFT protocol that relies on IP multicast, which can introduce interoperability problems, BFT-WS uses the regular SOAP/HTTP transport to avoid those issues.

Perpetual-WS\,\cite{4595891} also builds upon CLBFT and it attempts to address some of the shortcomings of Thema and BFT-WS, for instance, the support of replicated clients, or when a replicated Web Service has been compromised, which means it has more than \textit{f} faulty instances.
Perpetual-WS also supports long-running operations, as well as non-deterministic operations, such as local clock queries, pseudo-random numbers and timestamps, as the replicated Web Service will reach a consensus on the response to send back to the clients.
Albeit these advantages, the latency introduced by this middleware almost doubles, as the BFT algorithm is both run on the Web Service replicas as well as on the replicated clients, to reach consensus on the received responses, in order to avoid different responses to be accepted by them.
However, if the responding Web Service is compromised, all the clients might receive the same malicious answer and still agree to accept it, which really seems not to solve the indicated problem, contrarily to what was promised.

Similarly to Perpetual-WS, SWS\,\cite{Li:2005p3877} also enables the interaction between Web Services with different degrees of replication, and it additionally supports the dynamic discovery of Web Services by adding the replicas endpoints information to the service's WSDL, to allow UDDI registries to store and serve information on Web Services and their replicas.
However, it shares some of the shortcomings of Thema and BFT-WS.

A BFT algorithm that only requires message ordering per source instead of total ordering, thus reducing inter-replica communication, was proposed in\,\cite{6081845} to achieve trustworthy coordination of WS-BusinessActivity, allowing activities to tolerate the faulty behavior of the intervening parties.
Albeit having a better performance than other BFT protocols, this algorithm has limited application since it has been customized specially for WS-BusinessActivity, depending very closely on its state model. 

WS-FTM\,\cite{Looker:2005p3345} applies the N-Version model to Web Services in order to increase the dependability of a service, allowing it to tolerate both Byzantine and physical failures.
It uses a simple-majority voting scheme for achieving consensus on the response to a client's invocation, by analyzing the replies sent by the equivalent N-Version services in response to the replicated invocation. 

\section{Conclusion}
\label{sec:conclusion}

From the presented results, we can conclude that the presented implementation of the Raft algorithm is suitable for small scale scenarios, with a small number of clients and servers, performing better than Apache ZooKeeper on these settings by achieving better throughput and lower latency.
Moreover, the usage of DPWS provides a confident basis for the adoption of such a system to provide consensus in scenarios with largely heterogeneous devices, where churn could be overcome by the Ad-Hoc mode of WS-Discovery.
In scenarios with more concurrency, i.e. more clients and requests, Apache ZooKeeper clusters have the upper hand, expressed in the superior throughput which increases with the size of the cluster and seems to still haven't reached the saturation point, contrarily to Raft.

We point several possible directions for future work, in order to improve the work presented in this article.
An evaluation of how the presented results could be impacted by other types of failures, for instance on communications, and consequently triggered leader elections, as well as by the integration of advanced Raft features, such as changes to the cluster membership configuration, or log compaction, which allows leaders to send snapshots to followers instead of all the log entries, could be done.
The assessment of different strategies to accommodate distributed read requests issued by clients can also be performed, as well as the usage of different state machine implementations.
Further improvements on the implementation of the Raft protocol can be made, for instance, by using EXI, which can further improve the performance of this implementation, through the reduction of both the necessary time to process messages and the size of the exchanged messages\,\cite{Jammes:2014p21531}.

\section{Availability of Code}

The source code developed and used for experiments in this report is available as open source, allowing the experiments to be reproduced.
In detail, our implementation of Raft on the WS4D stack, that can be used to build applications, is available at \url{https://github.com/filipecampos/raft4ws}.
The code used for setting up and controlling the experiments described here is available at \url{https://github.com/filipecampos/raft_tests}.

\section*{Acknowledgment}
\small{
This work was made in the framework of the BEST CASE project (“NORTE-07-0124-FEDER-000056”) financed by the North Portugal Regional Operational Programme (ON.2 – O Novo Norte), under the National Strategic Reference Framework (NSRF), through the European Regional Development Fund (ERDF), and by national funds, through the Foundation for Science and Technology (FCT).
}

\bibliographystyle{abbrv}
\bibliography{article}

\end{document}